# A worldwide overview on the information security posture of online public services

João Marco Silva †
INESC TEC
University of Minho
joaomarco@di.uminho.pt

Diogo Ribeiro
INESC TEC
University of Minho
diogo.p.ribeiro@inesctec.pt

Luis Felipe Ramos
University of Minho
lfelipe.sm@gmail.com

Vítor Fonte ‡
UNU-EGOV
United Nations University
University of Minho
vitor.fonte@unu.edu

**Abstract**

*The availability of public services through online platforms has improved the coverage and efficiency of essential services provided to citizens worldwide. These services also promote transparency and foster citizen participation in government processes. However, the increased online presence also exposes sensitive data exchanged between citizens and service providers to a wider range of security threats. Therefore, ensuring the security and trustworthiness of online services is crucial to Electronic Government (EGOV) initiatives' success. Hence, this work assesses the security posture of online platforms hosted in 3068 governmental domain names, across all UN Member States, in three dimensions: support for secure communication protocols; the trustworthiness of their digital certificate chains; and services' exposure to known vulnerabilities. The results indicate that despite its rapid development, the public sector still falls short in adopting international standards and best security practices in services and infrastructure management. This reality poses significant risks to citizens and services across all regions and income levels.*

## 1. Introduction

Over the last decades, public administrations worldwide have been investing in Information and Communication Technologies (ICT) as a transformation strategy towards enhancing administrative processes, coverage, and efficiency of services provided to citizens. The modernisation effort driving Electronic Government (EGOV) also promotes government transparency and accountability (Alarabiat et al., 2018).

Biennially, the United Nations Department of Economic and Social Affairs (UN DESA) evaluates the EGOV development of all UN Member States by assessing several aspects of online services delivered by public administrations. The UN E-Government Survey encompasses whole-of-government approaches, open government data, e-participation, multi-channel service delivery, mobile services, usage uptake and digital divides, and innovative ICT-based partnerships (UN DESA, 2022).

Although revealing the global acceleration of public services' digitalisation – particularly during the COVID-19 pandemic – the UN DESA survey does not assess the information security and privacy aspects of the platforms supporting those services. Since sensitive data between citizens and EGOV service providers is transmitted through public communication infrastructures, adopting the best practices and technologies on the server side is crucial to the efforts of ensuring information security and trustworthiness in online public service delivery.

Therefore, this work provides the first worldwide overview of the security posture of EGOV providers regarding online platforms connecting services to citizens. The study evaluates 3068 unique online platforms from all UN Members States hosting public services of government ministries and federal agencies in three dimensions: (*i*) the support to up-to-date and robust protocols for secure end-to-end communications; (*ii*) the trustworthiness of the digital certificate chain of the provided services; and (*iii*) the exposure of the hosting servers to known vulnerabilities.

This analysis sheds light on the risks of having confidentiality, integrity, availability, and authenticity properties violated by widespread cyber threats due to deficient service configurations and inadequate patching and software update policies. Along with the discussion on the sources of security issues, the study also provides recommendations on how system administrators can improve services' security and overall information systems' robustness.

The next sections of this paper are organised as follows: Section 2 presents previous works addressing information security in EGOV services. Section 3 introduces the main technologies supporting the trust model of end-to-end secure communications, including digital certification chains. Section 4 presents the methodology used in this work to evaluate the security posture of EGOV service providers. Section 5 discusses the research outcomes, while Section 6 concludes this paper by providing recommendations for improving the information security of the surveyed platforms.

## 2. Related Work

In addition to adequate usability, coverage and comprehensiveness, the successful adoption of EGOV solutions also depends on the level of trust placed in them by citizens. Such trustworthiness is created through the perceived levels of security and privacy provided by the services and infrastructure hosting them. Typically, the details of implementation are not public, which hinders transparent analyses regarding the adoption of the best security practices and technologies by public service providers.

Regardless of this scenario, scientific literature has continuously addressed information security in the public sector over the last two decades. A common approach involves presenting and discussing policies, citizen engagement strategies, legal and normative frameworks, and the digital divide. Although they do not directly address information security issues and strategies, these works relate some of their analysis with the importance of investing in cybersecurity solutions and professionals (Ramos et al., 2021; Smith & Jamieson, 2006). Other works directly address security and privacy challenges in EGOV solutions within those topics (Shah et al., 2022; Yang et al., 2019).

Multiple authors have addressed how new technologies can be adopted in online public services and their impact on systems' and citizens' information security and privacy. Examples of these technologies are artificial intelligence and machine learning (Horowitz et al., 2018), cloud-based services (Susanto & Almunawar, 2016), and blockchain for government services (Alketbi et al., 2018).

Although less common, some works assess different aspects of information security in EGOV services and infrastructure deployments. However, they typically only focus on a single country or region, which does not provide a comprehensive overview of the global security posture adopted by the public sector. For instance, Alsmadi and Shanab (2016) and Ali and Zamri Murah (2018) performed penetration tests on platforms hosted on selected domain names of Jordan and Libya, respectively. Thompson et al. (2020) audited 40 websites from Australia and Thailand looking for security vulnerabilities, while Silva and Fonte (2019) analysed the Secure Sockets Layer and Transport Layer Security (SSL/TLS) protocols adoption in Portugal. A more comprehensive approach was adopted by Houser et al. (2022), who measured the availability and legitimacy of Domain Name Systems' (DNS) records in the authoritative *nameservers*, evaluating a database of domains obtained by DNS *lookup* rather than official domains of public services provided by the governments themselves.

Within this context, this work provides an unprecedented worldwide overview of the EGOV providers' practices and perceived posture toward hardening services and infrastructure through well-established technologies. It also assesses the level of exposure to multiple cyber-threats while providing recommendations to enhance the security of systems and citizens using them.

## 3. Background

Similarly to most online services, data between citizens and public service providers are typically transmitted over public networks, in which information security cannot be assured, and therefore, being exposed to a plethora of threats. In this scenario, two requirements are critical: guaranteeing the authenticity of the entity providing a service and protecting the confidentiality and integrity of transmitted data. Both requirements are addressed through protocols grounded in the field of cryptography, namely, digital certificates and public-key cryptography.

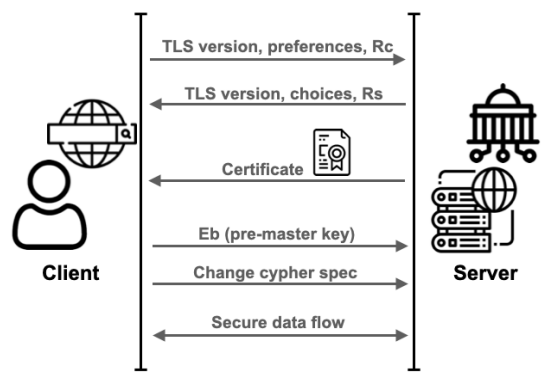

Figure 1: Simplified TLS handshake.

Figure 1 presents a simplified view of how the TLS protocol, currently in version 1.3 (Rescorla, 2018), implements these mechanisms in online communications. During the establishment of a secure session between the client device and the service server, there is a stage in which both entities exchange information about the cryptographic suites they support. They then negotiate cryptographic modes (and required parameters) and derivate the required cryptographic keys. Since this process relies on asymmetric cryptography and on key derivation protocols (*i.e.*, the symmetric key used to protect data is not transmitted), as long as the selected algorithms are strong, end-to-end confidentiality and integrity of data exchanged between parties can be assured even in the reasonable assumption of underlying unreliable and untrusted networks.

The authenticity of online services can be verified by resorting to digital certificates (SSL/TLS certificates for web-based applications), which consist of electronic documents or container files that contain a public key and identifying attributes about the entity that controls the associated private key (Whitman & Mattord, 2021). Certificates are issued by a trusted third party, *i.e.*, a *Certification Authority* (CA) and shared with a client during the TLS handshake (see Figure 1), which uses a Public Key Infrastructure (PKI) to validate the received certificate and verify the service identity. Globally, multiple CAs are organised hierarchically, meaning that verifying a certificate can require validating all the chain up to the top-level entity. A failure in this process precludes the trust placed on the authenticity of a certificate and, ultimately, on the security of an online platform.

Most browsers offer a simple way to verify the attributes of a session's certificate and issue alerts if the received certificate has been revoked, expired, or signed by an untrusted CA. This functionality also indicates that cryptographic protocols protect the communication between the user's device and the server within a session. However, the perception of security can be misleading since browsers do not evaluate the versions and robustness of the algorithms used to sign the certificate or encrypt the exchanged data.

Also, online platforms are often comprised of multiple publicly accessible software components, which translates to multiple open ports on public Internet Protocol (IP) addresses. It is not uncommon for multiple online platforms to share a common set of host servers. As a rule of thumb, complexity works against security and, the larger the potential attack surface (number of accessible components), the larger the number of known and unknown vulnerabilities, and the probability of a successful exploitation of one or more of them.

Thus, this work focuses on analysing these aspects in order to assess the actual trust citizens can reasonably place in the public services they interact with.

## 4. Methodology

To provide a worldwide overview of how the public sector addresses confidentiality, integrity and trustworthiness in online services, this work resorts to non-invasive scanning techniques to assess providers' security posture for communication between online services and citizens. This is particularly relevant since such communication mainly occurs via public networks, which are exposed to a myriad of threats.

A public dataset of 3068 unique domain names from all 193 UN Member States is used to identify and assess globally public platforms. This dataset includes domains hosting online platforms from government ministries, federal agencies, public service platforms, and the national portals provided to the United Nations E-Government Knowledge Database by national officials within the scope of the last UN E-Government Survey (UN DESA, 2022).

Table 1 presents the number of evaluated domains distributed according to their geographic regions and the countries' income following the World Bank's economy classification. Figure 2 demonstrates the domains' distribution in each region according to the countries' income. This detailing aims to put a geo-economic perspective on the analysis presented in Section 5.

Although Table 1 and Figure 2 only present five geographical regions, the analysis and discussion in Section 5 also take into consideration the 22 sub-regions used in the UN Geoscheme. In this regard, due to security and privacy concerns, this work does not reference specific domains or country names, and the analysis outcomes are presented as aggregates.

Table 1: Distribution of the analysed platforms.

| Region | | Income | |
|---:|---:|---|---:|
| Africa | 429 | High | 1233 |
| Americas | 269 | Upper-middle | 762 |
| Asia | 1261 | Lower-middle | 861 |
| Europe | 918 | Low | 212 |
| Oceania | 191 | | |
| Total | | 3068 | |

The assessment consists of inspecting the security posture of the online platforms hosted across all these domains, in three dimensions:

*1) Identifying whether the platforms support secure communications based on SSL/TLS protocols and analysing the cryptographic strength of the key exchange and the stream cyphers.* It is also analysed the adoption of more recent and secure protocols since older cryptographic protocol versions (*i.e.,* SSLv2,

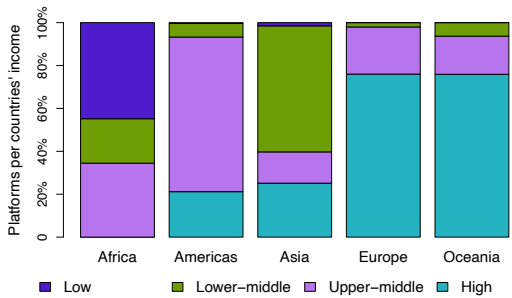

Figure 2: Platforms per country region and income.

SSLv3, TLS1.0, and TLS1.1) have known security vulnerabilities that can be exploited by attackers and are now deprecated by the Internet Engineering Task Force (IETF) (Housley, 2019). For platforms still supporting early versions, we inspect whether *weak* or *insecure* cyphers are active on their servers.

Notice that *weak cyphers* refer to cryptographic algorithms that are not broken or completely insecure but have unexploited known vulnerabilities, design flaws, or use key lengths regarded as too short to provide adequate protection against crypto-analysis techniques and attacks. On the other hand, *insecure cyphers* are those to which practical exploitation methods have been successfully developed.

This analysis phase is accomplished using *ssl-enum-cyphers*, a Network Mapper *(Nmap)*-based script (Nmap, 2023). *Nmap* is an open-source network scanning and security auditing tool designed to discover hosts and services on a computer network, thus providing a comprehensive map of the network. The specifically used script establishes multiple SSL/TLS connections, attempting different cyphers and data compressors and records whether the host accepts or rejects each of them. The qualitative analysis of servers' configuration is based on the SSL Server Rating Guide (Qualys, 2023b).

*2) Inspecting the chain of the SSL/TLS certificates used by the online services to assess whether the entire chain is properly configured, valid and trusted.* It includes evaluating the properties of the certificates, such as the strength and integrity of the signature algorithm, validity timeframe, and revocation status.

This analysis resorts to the *SSL Server Test* online service (Qualys, 2023c), which uses the SSL/TLS handshake to get the certificate chain of each surveyed platform, including the server certificate, as well as any intermediate and root certificates required to establish a chain of trust. Evaluating the entire certificate chain is vital to establish trust in communications between citizens and public services.

*3) In addition to assessing whether cryptographic protocols are exposed to known vulnerabilities (phase 1), all publicly accessible services running on the servers of online platforms were also assessed to establish the potential attack surface and actual known exploitable vulnerabilities in their software.*

This extended analysis utilises *Nmap* for probing open ports, identifying services and software version information. The returned Common Platform Enumeration (CPE) is then used to search known vulnerabilities in the public database for Common Vulnerabilities and Exposures (CVEs), maintained by the National Institute of Standards and Technology (NIST, 2023). The Common Vulnerability Scoring System (CVSS) is also used to evaluate exposure severity and risks.

The CVSS is an industry-standard framework used to assess and communicate the severity of computer systems and software security vulnerabilities. It provides a standardised and quantitative way of measuring the impact and exploitability of vulnerabilities, allowing organisations to prioritise and manage their response to security issues. In its latest version (3), vulnerabilities are scored as *low*, *medium*, *high* or *critical* according to their severity.

The reconnaissance scans adopted in this work are designed to ensure that information gathering does not access non-public or protected data, nor produce a significant load that could otherwise interfere with the regular operation of the platform.

## 5. Results

This section presents the results from the security analysis described in Section 4, organised into three groups: (*i*) the protocols supporting secure communications between public services and citizens; (*ii*) the trust levels provided by the deployed certificate chains; and (*iii*) the vulnerabilities to which the provided services are exposed to. The results correspond to the way the platforms were configured between December 2022 and January 2023.

### 5.1. Communication security

The survey of the online platforms hosted on the 3068 domains has demonstrated that TLS 1.2 continues to be the most widely deployed protocol version worldwide (see Figure 3). Although the support for SSL 2.0 and SSL 3.0 is residual (less than 2% each), TLS 1.0 and TLS 1.1 are still supported by 33.1% and 35.5% of the platforms, respectively, even considering that

IETF announced their deprecation in 2019 (Housley, 2019). This can potentially be explained by the need for backward compatibility, in order to support legacy systems or devices not compatible with newer TLS versions.

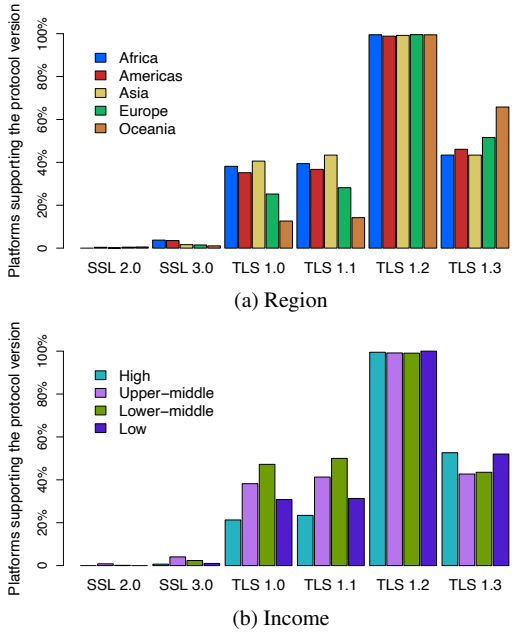

Figure 3: Protocol versions' distribution.

The most recent and secure TLS 1.3 specification was released in 2018 (Rescorla, 2018). However, during the time-frame of this study, less than 50% of the analysed platforms had added its support, which is significantly lower than the 62% estimated support of the global domains by May 2023 (Qualys, 2023a). This scenario might be related to the challenges of upgrading cryptography protocols (Ott et al., 2023).

Considering the distribution per region, Figure 3a shows the platforms hosted in domains from Oceania present a better relation between removing older versions and deploying TLS 1.3 (currently supported by over 65% of the domains from that region). Interestingly, Figure 3b shows that high-income and low-income countries have the same proportion of domains supporting TLS 1.3 (*i.e.,* around 52%), while upper-middle income countries are those with fewer domains supporting the newest protocol (*i.e.,* 42%). These results suggest that upgrading cryptographic suites depends more on governance practices than investment capacities. In fact, the most globally used software library for secure communications (*i.e.,* OpenSSL), which has supported TLS 1.3 since 2018, is distributed through open-source licensing (The OpenSSL Project, 2003), making it affordable to governments with budget constraints.

Although over 92% of the surveyed platforms support SSL/TLS-based connections, it does not mean that communications between citizens and public service servers are secure. Maintaining support for older and deprecated protocols exposes services and clients to high-risk threats. Namely, according to the National Vulnerabilities Database (NIST, 2023), the TLS 1.0 protocol is currently exposed to 163 vulnerabilities (18 with CVSS classified as *high*), while the TLS 1.1 protocol is currently exposed to 150 vulnerabilities (14 with a *high* CVSS classification).

Most vulnerabilities related to SSL/TLS communications result from *weak* and *insecure* cryptographic functions (*i.e.,* cyphers) supported by the protocol. For instance, TLS 1.0 and TLS 1.1 rely on *SHA-1 hash*, which can be used to perform *collision* attacks (Leurent & Peyrin, 2020). *MD5* algorithm also represents serious threats of *collision* and *preimage* attacks (Wang & Yu, 2005).

Therefore, by analysing the currently supported protocols, the study revealed that more than 77% of the evaluated platforms use at least one *weak* cypher while almost 5% still use at least one *insecure* cypher. Figure 4 details the results and shows that this scenario holds for all geographic regions and countries' income.

The sub-regions of Northern Africa, Melanesia and Southern Europe are those with the highest number of platforms with active *insecure* cyphers, *i.e.,* 16%, 12.5%, and 10%, respectively. Again, these results suggest that the level of income of a country does not correlate to the level of exposure of the cryptographic protocols being used since over 98% of the European online platforms are hosted in upper-middle or high-income countries (see Figure 2).

When considering the number of *weak* and *insecure* cyphers active in platforms of each domain, Figure 5 and Figure 6 show an even more concerning scenario. Globally, over 1500 platforms have over 10 active *weak* cyphers (see Figure 5). Despite the lower numbers compared to the *weak* cyphers, the global median of *insecure* cyphers per domain, being 2, represents a higher risk (see Figure 6). Moreover, Figure 6b shows the distribution of *insecure* cyphers is similar across all incomes (with outlier cases among the upper-middle and lower-middle income countries).

## 5.2. Certificate chain trustworthiness

As introduced in Section 3, digital certificates underpin the security model in which citizens can trust the services they are communicating with are actually authentic. To do so, besides providing SSL/TLS certificates to all services, system administrators must ensure a trusted CA signs them and that all the chain from the local certificate to the root certificate is valid (non-revoked) and use secure signatures and key sizes.

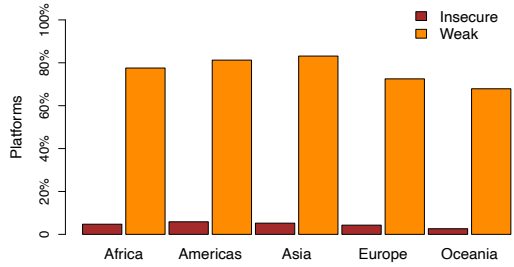

(a) Region

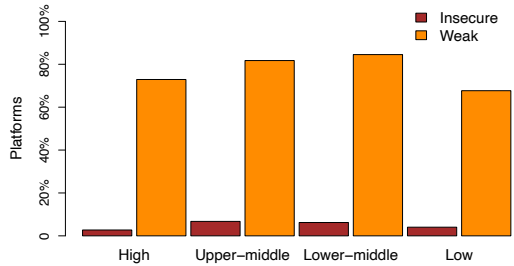

(b) Income

Figure 4: Distribution of *weak* and *insecure* cyphers.

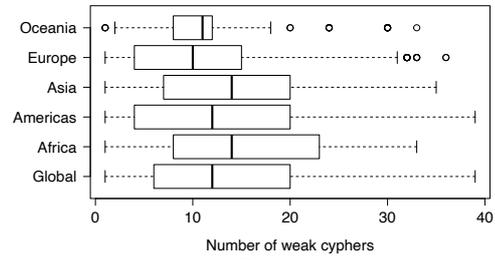

(a) Region

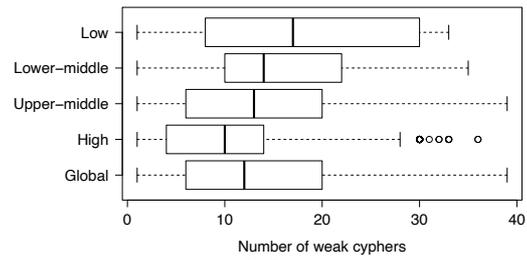

(b) Income

Figure 5: Distribution of *weak* cyphers per platform.

Despite the fundamental role in providing trust to citizens, Figure 7 shows the alarming reality that: over 80% of the domains hosting the analysed platforms present at least one issue related to their SSL/TLS certificate chains; over 8% are signed by an untrusted CA; or present issues in their root certificates. Although with small variations, this scenario is observed in domains from all regions (see Figure 7a). Considering sub-regions, certificate issues are present in over 90% of the analysed domains from countries in Western Africa, the Caribbean, Southern Asia, and Micronesia.

When considering country incomes, Figure 7b shows that over 96% of the domains from the low-income tier have certificate-related issues. Emphasising these results' relevance, it should be noted that government representatives shared the analysed domains, stating them as hosting strategic national e-government platforms (UN DESA, 2022).

A deeper analysis of these results reveals the most common issues in certificates used by the surveyed platforms. They are depicted in Figure 8 along with the underlying terminology. Globally, over 68% of the analysed platforms use *insecure signatures*, meaning either the private key used to sign the certificate or the hashing function used in the signature is *weak* or *insecure*. The analysis also revealed that the insecure *SHA-1 hashing function* is the primary source of such issues.

Another source of certificate-related issues is *host name mismatches*. This happens when the *Common Name* attribute of a certificate does not match the URL used to reach the service, leading to authenticity threats since users cannot verify the server identity. More than 70% of low-income countries present such an issue (see Figure 8b). The same can be observed in over 57% of the analysed African online platforms (see Figure 8a).

Also, more than 27% of the evaluated platforms did not have a valid certificate by the time of this analysis due to expiration. This is represented in Figure 8 by the issue *not after* (NAF), meaning the certificate in use had this attribute set to a date previous to this study. Again, this issue is more common in low-income countries (see Figure 8b) and might be related to the costs of issuing new certificates. Domains from Oceania and Africa are also the ones with more incidence of expired certificates in use (i.e., 41.58% and 39.15%, respectively).

A positive discovery about digital certificates used by the analysed public services is the overall low number of *revoked* (RVK) and *blocklisted*[1] (BLK) certificates,

---

[1] This term was adopted in order to comply with ACM's "Words

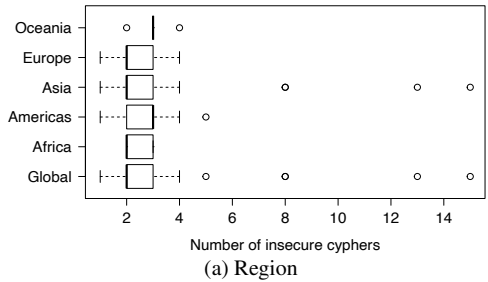

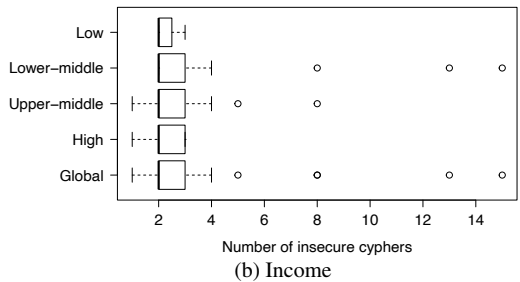

Figure 6: Distribution of *insecure* cyphers per platform.

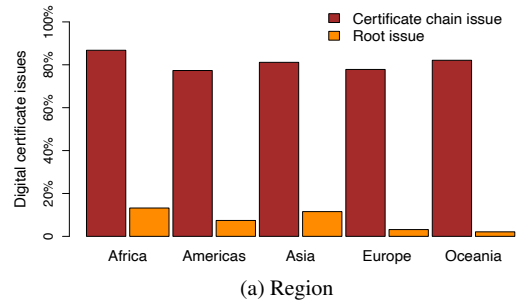

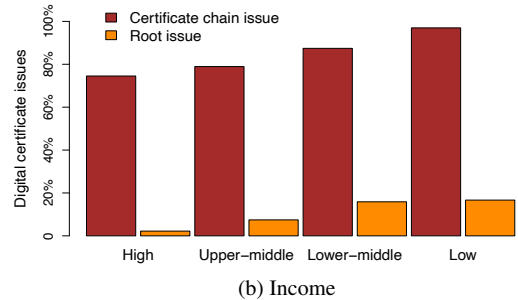

Figure 7: Issues in the SSL/TLS certificate chain.

also depicted in Figure 8.

### 5.3. Software vulnerability

The third dimension of this study provides an overview of known vulnerabilities identified in the software running on publicly accessible hosting servers of the online platforms. They typically stem from using software components with security flaws or weaknesses that increase attack surfaces and expose both services and citizens to multivector threats.

Figure 9 presents the proportion of the domains with at least one known vulnerability, considering region and income groups. Globally, over 20% of the analysed domains were exposed to at least one vulnerability by the time of this analysis. Here, economic realities seem to represent a causal factor since over 45% of the low-income countries were exposed, contrasting to around 9% of the high-income countries.

Geographically, over 33% of the services from African domains were exposed to at least one known vulnerability. Considering the sub-regions, Middle Africa and Micronesia domains were more exposed (*i.e.,* over 50% of vulnerable services).

The domains not exposed to any known vulnerability (*i.e.,* approximately 80%) might suggest worldwide

---

Matter" alternative terminology recommendations, available at https: //www.acm.org/diversity-inclusion/words-matter.

solid practices regarding infrastructure protection. However, the dynamics concerning the discovery and publication of vulnerabilities require continuous monitoring and response by platform administrators, which can be challenging to low-income and lower-middle-income countries.

Considering the number of vulnerabilities per domain, Figure 10 depicts the distribution per region (*i.e.,* Figure 10a) and per countries' income (*i.e.,* Figure 10b). At the time of this analysis, 72 domains were exposed to more than 100 known vulnerabilities, mostly from low-income countries.

Although this study does not provide details regarding criticality of the exposure of specific online platforms, the current scenario can be justifiably perceived as of a significant threat. More specifically, the study has identified 767 different vulnerabilities across the surveyed domains, 63 of them with *severity* classified as *critical*, according to the CVSS v.3, and 341 classified as *high*. These vulnerabilities were identified in 11529 different services, meaning that such systems have a higher risk of being the target of damaging attacks.

Table 2 presents the top 20 vulnerabilities regarding severity and number of affected online platforms. It is important to highlight that 12 vulnerabilities are classified with *critical* or *high* severity. Moreover, half of them were discovered and published over one year

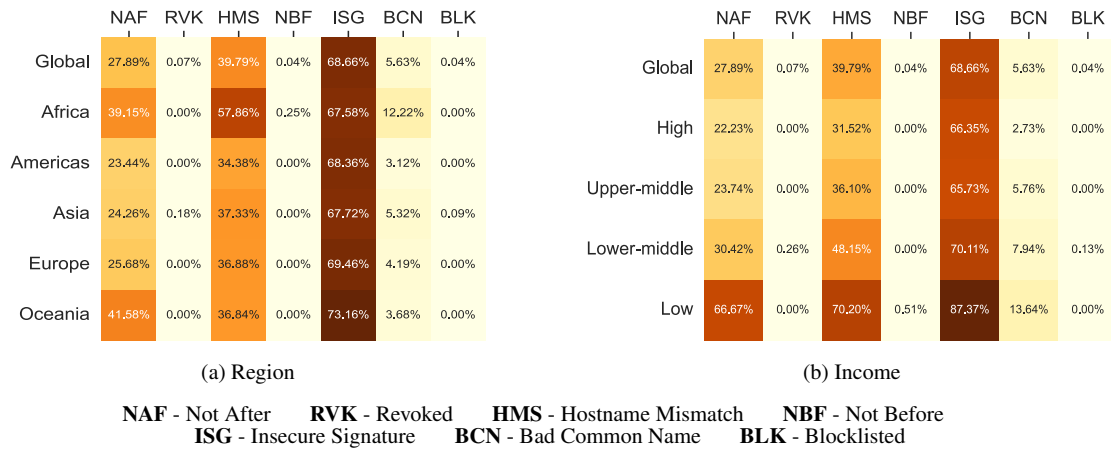

(a) Region  (b) Income

**NAF** - Not After  **RVK** - Revoked  **HMS** - Hostname Mismatch  **NBF** - Not Before
**ISG** - Insecure Signature  **BCN** - Bad Common Name  **BLK** - Blocklisted

Figure 8: Distribution of certificate problems.

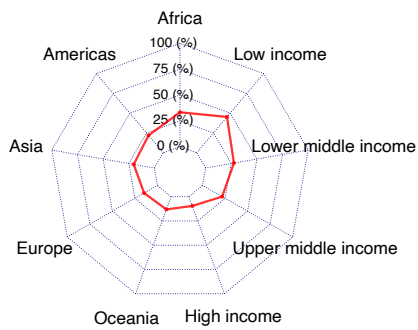

Figure 9: Distribution of the exposed platforms.

before this study, which suggests that several platforms do not have a monitoring and response policy for new known vulnerabilities. Combined with the severity assessment stated above, this further increases security risks for both infrastructures and citizens.

## 6. Discussion and Conclusions

This study provides a comprehensive overview of the security posture of EGOV providers worldwide regarding online service delivery. By evaluating online platforms hosted across 3068 unique domains provided by national officials of all UN Member States, encompassing government ministries, federal agencies, public service platforms, and their national portals, the analysis focuses on assessing their support to secure communication protocols, the trustworthiness of their digital certificate chain, and services exposure to known vulnerabilities.

The findings reveal both progress and areas of concern. While TLS 1.2 remains the most widely deployed protocol for secure communications, there is still significant support for older versions, such as TLS 1.0 and TLS 1.1, despite their deprecation. Although recognising scenarios requiring the support of earlier versions for compatibility purposes, it is recommended to phase out *weak* and *insecure* cyphers from the underlying cryptography suites. Even with open-source libraries offering reduced-cost upgrades, it was observed a relatively lower adoption of the latest and more secure TLS 1.3. Some factors that may influence the upgrading pace include a lack of skilled professionals and financial constraints to support the costs of acquisition and constant maintenance of secure ICT infrastructure across public service providers of countries from all income tiers (Glyptis et al., 2020; Ramos et al., 2021). This scenario evinces the requirements for long-term investments in technical training and threat awareness of systems administrators.

The assessment of certificate chains highlights the need for proper configuration, validity, and trust. Ensuring the integrity of the entire chain is crucial for establishing trust in communications between citizens and public services. Since the main issue identified is the use of the insecure *SHA-1 hashing function*, a straightforward action is to follow the major certificate issuers' practices and adopt the more secure *SHA256*. In addition to adopting the best practices in certificate management, such as avoiding self-signed certificates, properly securing private keys, ensuring timely renewal, and revoking compromised certificates, public administration must foster citizens' security literacy. Endowing citizens with the knowledge to assess the trustworthiness of the services they interact with can reduce the risks of cyber attacks.

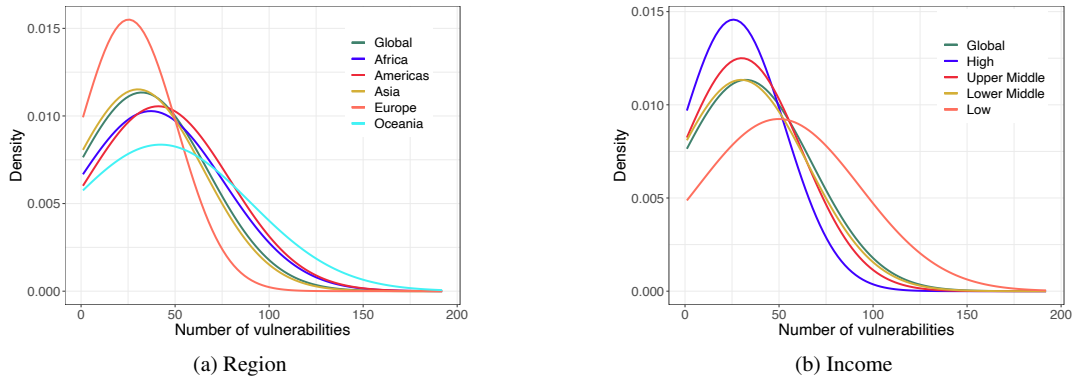

(a) Region　　　　　　　　　　　　(b) Income

Figure 10: Number of vulnerabilities per platform.

Table 2: Top 20 vulnerabilities to which platforms are exposed - Global overview.

| CVE | Severity | Domains exposed | CVE | Severity | Domains exposed |
|---|---|---|---|---|---|
| CVE-2020-14145 | Medium | 252 | CVE-2022-22719 | High | 177 |
| CVE-2021-41617 | High | 250 | CVE-2022-22720 | Critical | 177 |
| CVE-2018-15919 | Medium | 183 | CVE-2022-22721 | Critical | 177 |
| CVE-2022-28614 | Medium | 181 | CVE-2019-6111 | Medium | 176 |
| CVE-2022-28615 | Critical | 181 | CVE-2019-6110 | Medium | 176 |
| CVE-2022-26377 | High | 181 | CVE-2018-20685 | Medium | 176 |
| CVE-2022-29404 | High | 181 | CVE-2019-6109 | Medium | 176 |
| CVE-2022-30556 | High | 181 | CVE-2021-44790 | Critical | 169 |
| CVE-2022-31813 | Critical | 181 | CVE-2019-17567 | Medium | 166 |
| CVE-2022-23943 | Critical | 177 | CVE-2021-26690 | High | 166 |

The study also identifies vulnerabilities in the hosting servers of EGOV providers, emphasising the risks of confidentiality, integrity, availability, and authenticity violations. Deficient service configurations and inadequate patching and updated policies contribute to these risks. Therefore, in addition to addressing vulnerabilities through timely patching and updates, it is crucial that administrators adopt regular security audits, deploy automatic vulnerability assessments, perform regular penetration testing, secure network devices, adopt industry standards for incident response, continuously monitor the supporting infrastructures, and adopt threat intelligence.

By shedding light on the security challenges faced by EGOV providers, this study contributes to the ongoing efforts to enhance the trustworthiness and security of digital government services. It does so by advocating that investment decisions related to ICT must also take into account the costs of preventing and responding to security incidents, which, in the medium to long term, would otherwise translate into damaging security-related incidents, hampering public trust in online service-delivery, and incurring in high costs of remediation when compared to maintaining up-to-date cryptographic protocols and software.

**Acknowledgements:** † This work is financed by National Funds through the Portuguese funding agency, FCT - Fundação para a Ciência e a Tecnologia, within project LA/P/0063/2020. ‡ This research was partially supported by project "INOV.EGOV-Digital Governance Innovation for Inclusive, Resilient and Sustainable Societies / NORTE-01-0145-FEDER-000087", funded by the NortePortugal Regional Operational Programme (NORTE 2020), under the PORTUGAL 2020 Partnership Agreement, through the European Regional Development Fund (EFDR).